# High-Temperature Strong Nonreciprocal Thermal Radiation from Semiconductors

Bardia Nabavi[1], Sina Jafari Ghalekohneh[1], Komron J. Shayegan[2,3], Eric J. Tervo[4,5], Harry Atwater[2,3], and Bo Zhao[1,*]

[1]Department of Mechanical and Aerospace Engineering, University of Houston, Houston, TX 77204, USA

[2]Thomas J. Watson Laboratory of Applied Physics, California Institute of Technology, Pasadena, CA, 91125, USA

[3]Department of Electrical Engineering, California Institute of Technology, Pasadena, CA, 91125, USA

[4]Department of Electrical and Computer Engineering, University of Wisconsin-Madison, Madison, WI, 53706, USA

[5] Department of Mechanical Engineering, University of Wisconsin-Madison, Madison, WI, 53706, USA

*Corresponding author: bzhao8@uh.edu

**Abstract**

Nonreciprocal thermal emitters that break the conventional Kirchhoff's law allow independent control of emissivity and absorptivity and promise exciting new functionalities in controlling heat flow for thermal and energy applications. In enabling some of these applications, nonreciprocal thermal emitters will unavoidably need to serve as hot emitters. Leveraging magneto-optical effects, degenerate semiconductors have been demonstrated as a promising optical material platform for nonreciprocal thermal radiation. However, existing modeling and experimental efforts are limited to near room temperature (< 373 K), and it remains unclear whether nonreciprocal properties can persist at high temperatures. In this work, we demonstrate strong nonreciprocal radiative properties at temperatures up to 600 K. We propose a theoretical model by considering the temperature dependence of the key parameters for the nonreciprocal behavior and experimentally investigate the temperature dependence of the nonreciprocal properties of sufficiently doped InAs, a degenerate semiconductor, using a customized angle-resolved high-temperature magnetic emissometry setup. Our theoretical model and experimental results show agreement, revealing that strong nonreciprocity can persist at temperatures over 800 K for high-temperature stable semiconductors, enabling a pathway for nonreciprocal radiative heat flow control at high temperatures.

## Introduction

Recent decades have witnessed remarkable advances in thermal radiation control through engineered nanophotonic structures.[1,2] Beyond controlling thermal radiation with subwavelength structuring, integration of nonreciprocal materials into photonically engineered thermal emitters has emerged as a new route to unequal emissivity and absorptivity properties, due to the breaking of Lorentz reciprocity.[2-5] Through magneto-optic effects[2-5], nonlinear responses[6], or dynamic time modulation[7,8], nonreciprocal systems enable independent control of emissivity and absorptivity. The ability to control radiative properties promises novel functionalities for heat flow control[5,9] and energy harvesting.[10-12]

Using magneto-optical materials among other approaches to achieve nonreciprocity has been explored in photonics for many years.[13] These materials are typically semiconductors that support nonreciprocal response induced by the cyclotron motion of free electrons.[14] Since early theoretical studies, advances have been made in achieving nonreciprocal thermal radiative properties leveraging the nonreciprocal responses in these materials.[1,5] For example, Shayegan et al.[15,16] experimentally demonstrated the inequality between emissivity and absorptivity of *n*-doped indium arsenide (InAs) planar and grating structures. Liu et al.[17] achieved highly asymmetric nonreciprocal absorption across a broadband long-wave infrared range (20-35 μm) using fourteen InAs layers with gradient doping concentrations corresponding to cascaded epsilon-near-zero (ENZ) frequencies. A similar effect can be observed in other semiconductor and magnetic Weyl semimetal systems, as discussed in Refs. 17 and 18. Recently, Do et al.[18] used Bayesian optimization and designed a multilayer InAs and magnetic Weyl semimetal hybrid structure with only a few layers that can achieve strong nonreciprocal emission from 5 to 40 μm, highlighting the great potential for semiconductors in achieving nonreciprocal thermal emission.

In enabling functionalities associated with radiative heat flow control, nonreciprocal thermal emitters that can sustain high temperatures are critical.[11] It is expected that semiconductor properties, such as free carrier concentration, exhibit a strong temperature dependence.[15,19] While significant progress has been made in experiments, most studies have focused on near-room temperatures, with experimental conditions rarely exceeding 423 K[15], below which the intrinsic carrier concentration is negligible compared to extrinsic carriers in degenerate samples.[19,20] As temperatures go well above room temperature, the change in carrier concentration[19], effective electron mass[21,22], high-frequency permittivity[21], and electron mobility[22] becomes non-negligible. However, it is not clear how these material property changes could impact the nonreciprocal properties.

In this work, we investigate the high-temperature nonreciprocal thermal radiative properties from semiconductors. We propose a model for the dielectric functions of the semiconductors by considering the temperature dependence of material parameters relevant to nonreciprocity. As a verification of the model, we measure the temperature-dependent, nonreciprocal radiative properties of two InAs samples with different doping using an in-house developed, angle-resolved, high-temperature magnetic emissometry setup. We show that, despite a decrease in nonreciprocal contrast between emissivity and absorptivity, a strong nonreciprocal effect persists even at temperatures over 600 K.

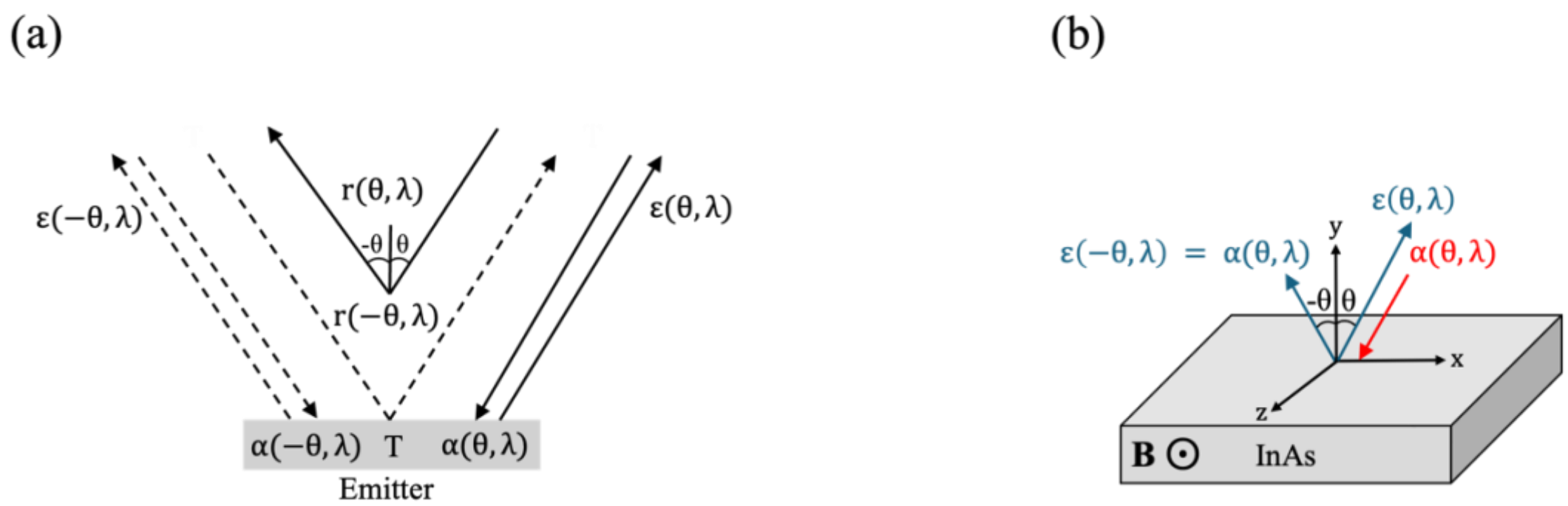

Figure 1. (a) Schematic of radiative heat transfer paths for a thermal emitter with specular reflection. Solid and dashed arrows show the energy flow balance for beams incident from the $+\theta$ or $-\theta$ directions, respectively. (b) Illustration of the InAs structure considered in this work.

## Theoretical Model

The nonreciprocal thermal radiative properties are characterized by the imbalance between the directional spectral emissivity ($\varepsilon$) and absorptivity ($\alpha$). These properties can be computed from scattering properties such as reflectivity ($r$). Here, we consider opaque planar nonreciprocal thermal emitters, as depicted in Fig. 1(b). These emitters possess a specular reflection, and the emissivity and absorptivity can be obtained as:[23,24]

$$\varepsilon(\theta,\lambda) = 1 - r(-\theta,\lambda) \tag{1}$$

$$\alpha(-\theta,\lambda) = 1 - r(\theta,\lambda) \tag{2}$$

where $r$ can be computed by solving Maxwell's equations at the interface. In this study, we use an in-house developed nonreciprocal rigorous coupled-wave analysis algorithm.[25] The level of nonreciprocity in radiative properties can be evaluated using the contrast between absorptivity and emissivity, expressed as:

$$\eta(\theta,\lambda) = |\alpha(\theta,\lambda) - \varepsilon(\theta,\lambda)| \tag{3}$$

The nonreciprocal radiative properties result from the nonreciprocal dielectric tensor by the magneto-optical behavior of semiconductors in a magnetic field. In general, when the magnetic field (**B**-field) is applied along the *z* direction, the temperature-dependent dielectric tensor can be described with a Drude model as:[4,26,27]

$$\varepsilon = \begin{bmatrix} \varepsilon_{xx} & \varepsilon_{xy} & 0 \\ \varepsilon_{yx} & \varepsilon_{yy} & 0 \\ 0 & 0 & \varepsilon_{zz} \end{bmatrix}, \tag{4}$$

$$\varepsilon_{xx} = \varepsilon_{yy} = \varepsilon_{\infty} - \frac{\omega_{\mathrm{p}}^2(T)(\omega+\mathrm{i}\gamma(T))}{\omega[(\omega+\mathrm{i}\gamma(T))^2-\omega_{\mathrm{c}}^2(T)]}, \tag{5}$$

$$\varepsilon_{xy} = -\varepsilon_{yx} = \mathrm{i}\frac{\omega_{\mathrm{p}}^2(T)\omega_{\mathrm{c}}(T)}{\omega[\left(\omega+\mathrm{i}\gamma(T)\right)^2-\omega_{\mathrm{c}}^2(T)]}, \tag{6}$$

$$\varepsilon_{zz} = \varepsilon_{\infty} - \frac{\omega_{\mathrm{p}}^2(T)}{\omega(\omega+\mathrm{i}\gamma(T))} \tag{7}$$

where $\varepsilon_{\infty}$ is high-frequency relative permittivity, $\gamma$ is electron scattering rate, $\omega$ is angular frequency, $\omega_{\mathrm{c}}$ is cyclotron frequency defined as $\omega_c = eB/m^*$, where *e* is the elementary charge, *B* is magnetic field density, *T* is temperature, and $m^*$ is effective electron mass.[4,26,27] $\omega_{\mathrm{p}} = \sqrt{ne^2/m^*\varepsilon_0}$ is plasma frequency where $n$ is free carrier concentration and $\varepsilon_0$ is vacuum permittivity.[4,26,27] Most of the parameters, as indicated in the above equations, are prone to change with varying temperature, $T$.[27] The key to modeling the nonreciprocal properties, therefore, relies on capturing the temperature dependence of these parameters. Here, we use *n*-doped InAs as shown in Fig. 1(b) as an example to illustrate our model. With the dielectric tensor discussed above, such a system supports nonreciprocal responses for the transverse magnetic (TM) polarization (with the magnetic field along the *z*-direction). Other semiconductors can be modeled using a similar procedure using different parameters or equations for the material[28-30], and we detail the process in the Supplemental Materials.

The plasma frequency, which approximately determines the frequency of large nonreciprocal behavior, depends on both the free carrier concentration and the electron effective mass. The temperature dependence of the free carrier concentration is mainly caused by changes in band gap and effective density of states of both the conduction and valence bands:[19,31]

$$E_{\mathrm{g}} = E_{\mathrm{g0}} - \frac{aT^2}{T+b}, \quad (8)$$

$$N_{\mathrm{c}} = 1.68 \times 10^{13} T^{1.5}, \quad (9)$$

$$N_{\mathrm{v}} = 1.27 \times 10^{15} T^{1.5}, \quad (10)$$

$$n_{\mathrm{i}} = \sqrt{N_{\mathrm{c}} N_{\mathrm{v}}} e^{-\frac{E_g}{2kT}}, \quad (11)$$

$$n = N_{\mathrm{D}} + \frac{n_{\mathrm{i}}^2}{N_{\mathrm{D}}} \quad (12)$$

where $E_{\mathrm{g}}$ is the band gap, $E_{\mathrm{g0}}$, $a$, and $b$ are 0.415 eV, $2.76\times10^{-4}$ eV/K, and 83 K, respectively.[31,32] $N_{\mathrm{c}}$ and $N_{\mathrm{v}}$ are effective densities of states in the conduction and valence bands, respectively, $k$ is the Boltzmann constant, $n_{\mathrm{i}}$ is the intrinsic carrier concentration, $N_{\mathrm{D}}$ is doping level, and $T$ is the absolute temperature. Equations (9) and (10) capture the main effect of a changing temperature on $N_{\mathrm{c}}$ and $N_{\mathrm{v}}$,[33] with the higher-order effect of changing effective mass neglected. Furthermore, we note that the temperature effect on carrier concentration is very mild for temperatures below 500 K for n-doped samples with a doping level lower than $10^{16}$ $\mathrm{cm}^{-3}$.[19] In this region, which is called the extrinsic-region, the majority of free carriers come from dopants and the thermally excited carriers are negligible.[19] The change in free carrier concentration impacts the dielectric function negligibly in this regime, which might seem counterintuitive.

The temperature threshold up to which the impact on the carrier concentration of the semiconductor can be neglected depends largely on the band gap and the doping level.[19] For larger band gap semiconductors, for example, gallium arsenide (GaAs), the carrier concentration does not change significantly up to 600 K for doping levels of $10^{18}$ cm$^{-3}$, as shown in the Supplemental Materials.

The effective electron mass is influenced by temperature through its dependence on the band gap:[34,35]

$$\frac{m^*}{m_\mathrm{e}} = 0.0310 - 0.0218E_\mathrm{g} + 0.1E_\mathrm{g}^2 - 0.046E_\mathrm{g}^3 \qquad (13)$$

where $m_\mathrm{e}$ is the free electron mass. Decreasing band gap with increasing temperature reduces the effective mass at higher temperatures, thereby increasing the cyclotron frequency for a given magnetic field. Therefore, the change in effective mass contributes a stronger nonreciprocal effect at higher temperature.

Additionally, the scattering rate varies with temperature. We determine the temperature effect on the electron scattering rate using a widely used empirical model for low-field mobility:[22]

$$\mu_\mathrm{LF}(n,T) = \mu_\mathrm{min} + \frac{\mu_\mathrm{max}\left(\frac{380}{T}\right)^{\psi_1} - \mu_\mathrm{min}}{1+\left(\frac{n}{n_\mathrm{ref}\left(\frac{T}{300}\right)^{\psi_2}}\right)^{\zeta}} \qquad (14)$$

where $n_\mathrm{ref}$ is reference doping concentration $1.1 \times 10^{18}$cm$^{-3}$, $\mu_\mathrm{max}$ is mobility at 300 K, and $\mu_\mathrm{min}$ is the minimum mobility. The respective mobility values are 34000 and 1000 cm$^2$/V·s for *n*-doped InAs.[22] The experimentally-determined $\psi_1$, $\psi_2$, and $\zeta$ [22] are 1.57, 3, and 0.32, respectively. The electron scattering rate can be derived from the electron mobility as follows:[33]

$$\gamma = \frac{q}{\mu_\mathrm{LF} m^*} \qquad (15)$$

where $q$ is the elementary charge. As shown in the supplementary materials, an increase in temperature causes an increase in electron scattering.

Besides the parameters describing free electrons in a semiconductor, previous research indicate that the high-frequency relative permittivity $\varepsilon_\infty$ also varies with temperature.[21] The importance of this dependence has been noted in Ref. 15, and it directly impacts the spectral location of the ENZ region at different temperatures. Based on the tabulated data from Ref. 21, we derive a linear model using least-squares regression to capture this dependence:

$$\varepsilon_\infty = \delta T + \sigma \tag{16}$$

Here, $\delta$ and $\sigma$ are 0.0018 $K^{-1}$ and 13.54, respectively. The positive high-frequency permittivity competes with the negative Drude term, which determines the spectral location of the ENZ region. As temperature increases, the high-frequency permittivity increases, causing a redshift of the ENZ region.

The overall temperature dependence arises from the collective contributions of all the above-mentioned parameters. Among these, the scattering rate and high frequency permittivity cause significant changes in permittivity as the temperature varies, which is confirmed by calculating the permittivity of the material at high temperatures while keeping all other parameters constant. In the Supplementary Materials, we present the temperature dependence of all parameters for two representative magneto-optical materials: InAs and GaAs and their stoichiometric combination, $In_{0.53}Ga_{0.47}As$. Figure 2 illustrates the temperature dependence of the diagonal dielectric function for InAs based on our model. As temperature increase, a notable redshift of the ENZ region is observed, driven by the simultaneous increase in real and imaginary parts of the permittivity.

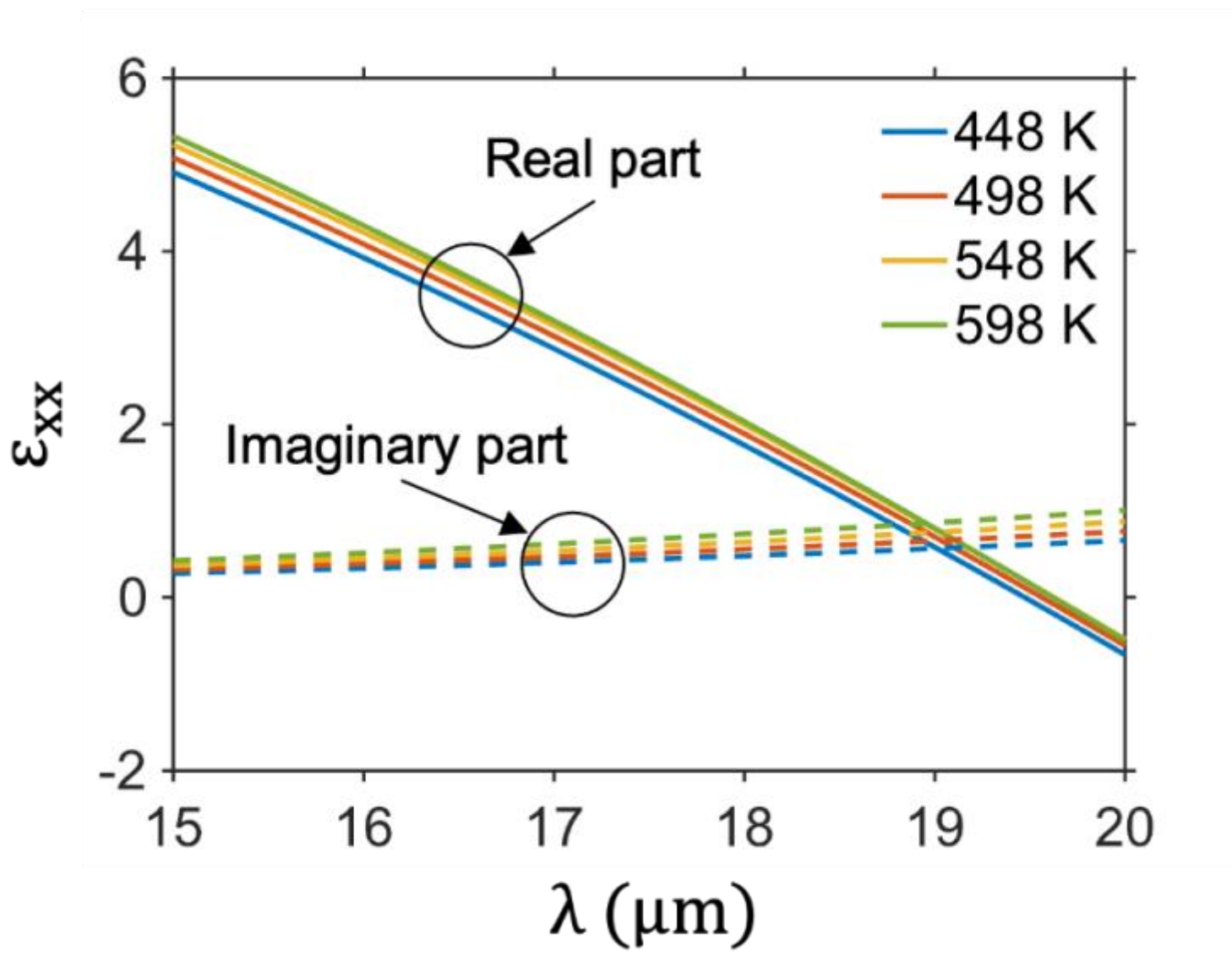

Figure 2. $\varepsilon_{xx}$ as a function of wavelength $\lambda$ for $n$-doped InAs with doping level of $1.4 \times 10^{18}$ cm$^{-3}$ at different temperatures.

**Experimental Measurement Setup**

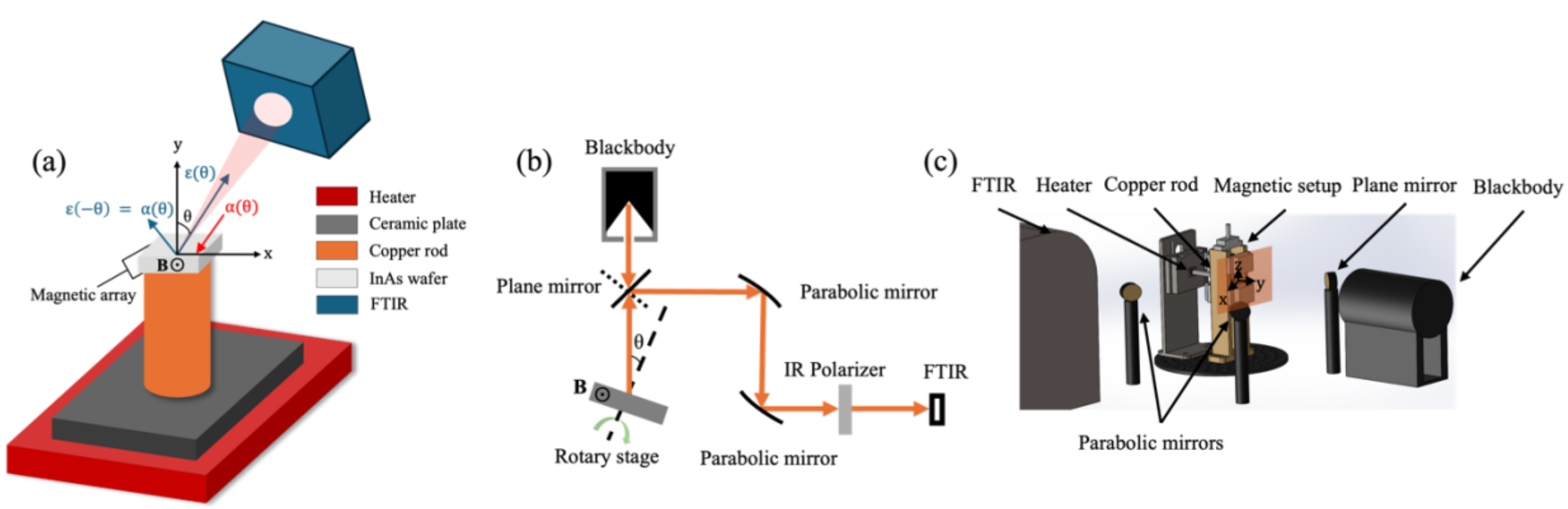

Figure 3. Schematic of the emissometry setup. (a) A ceramic heater ensures uniform heating. A copper rod guides the heat to the sample, which is attached to the end of the rod. An array of magnets surrounds the sample, providing the external magnetic field. The emission is detected by the FTIR to measure the emissivity of the samples. (b) The beam path of the emissometry setup. (c) Schematic of the emissometry setup showing the arrangement of all parts.

We conduct an experimental demonstration of the radiative properties of single crystal *n*-doped InAs samples as a verification of our model. We develop a high-temperature angle-resolved magnetic emissometry setup for this purpose, as shown in Fig. 3. Samples are mounted on a sample stage, which is a PID-controlled ceramic heater (Instec Inc. model number HS1200G). The sample stage is placed on a freely rotatable plate, allowing for changes in the emission direction. The emitted radiation from the sample is directed through a plane gold mirror and two parabolic gold mirrors to a Fourier Transform Infrared (FTIR) spectroscopy system, where a zinc selenide (ZnSe) polarizer can be added to the beam path. The plane mirror can be reconfigured to guide the emission from a blackbody source, enabling spectral measurements of the reference blackbody at the same temperature as the sample. The spectral directional emissivity of the sample can then be measured as:[36]

$$\varepsilon(\lambda, \theta, B, T_s) = \varepsilon_R(\lambda) \frac{I_S(\lambda,\theta,B,T_s) - I_{BG}(\lambda,\theta,T_s)}{I_R(\lambda,T_s) - I_{BG}(\lambda,\theta,T_s)} \tag{17}$$

Where $B$ is magnetic field density, and $T_S$ is temperature of the sample. $I_S$, $I_R$, and $I_{BG}$ refer to measured spectral intensity from the sample, the reference blackbody, and the background, respectively. The emissivity of the reference material, $\varepsilon_R$, is the emissivity of a blackbody and equal to one for all wavelengths in our measurements. A thick copper foil is used to cover the sample stage to block the radiation from the other parts of the sample stage, reducing the background emission as much as possible. To quantify the remaining spectral intensity of the background, a thin aluminum foil of the same area as the sample is attached at the tip of the rod to replace the sample, and the intensity is recorded at the same temperature as the InAs samples. From Eqs. (1) and (2), one can see that $\varepsilon(\theta, \lambda) = \alpha(-\theta, \lambda)$. Therefore, the absorptivity at $-\theta$ can be obtained from the emissivity at angle $\theta$, providing an approach to obtain the absorptivity directly from the emission measurement. We use this approach in our experiment to obtain the absorptivity, emissivity, and the nonreciprocal contrast of InAs.

To induce the nonreciprocal response, a magnetic field is required. Here, the external magnetic field is applied to the sample via a magnetic assembly placed on the rotary stage. The design of the magnetic setup is shown in Fig. 4. Inspired by[37], we place four pairs of neodymium (NdFeB) magnets to enhance the magnetic field strength between two supermendur rods. In doing so, a **B**-field up to 1 T can be achieved in the 1 cm gap between the two supermendur rods, where the sample can be placed.

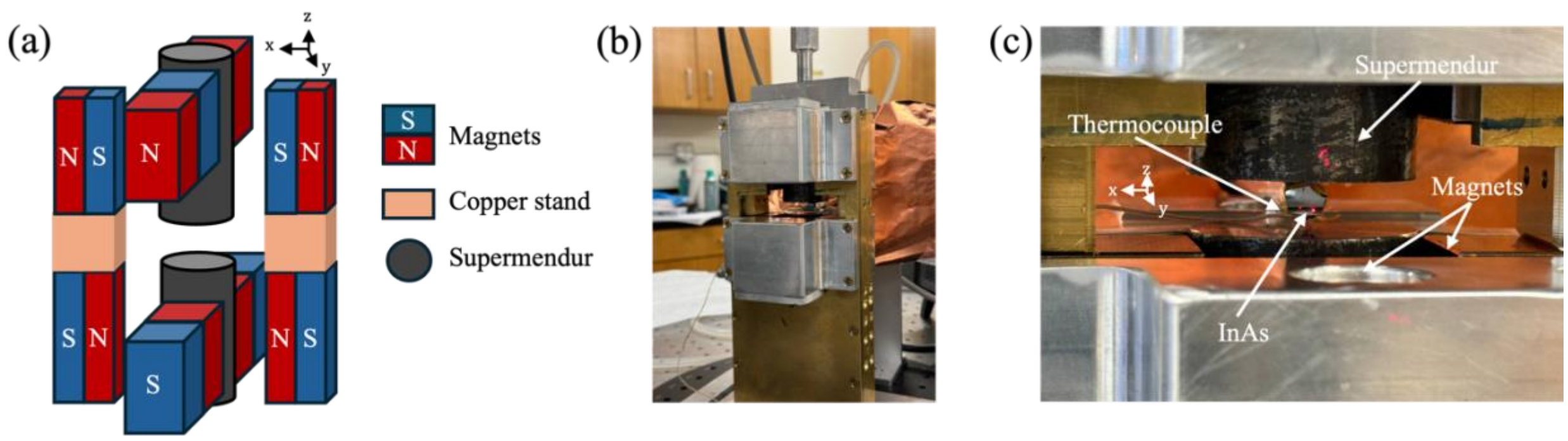

Figure 4. Schematic of the magnet assembly. (a) N and S represent the north and south poles of the permanent magnets. Eight neodymium magnets are arranged surrounding the supermendur, providing a magnetic field as high as 1 T. Magnets and supermendur are mounted on a copper stand. (b) Picture of the copper stand of the magnetic assembly and stainless-steel fixture. (c) Picture of the sample in the magnetic field. Thick copper foil with a round hole for the fin structure is used to block radiation coming from the sample stage.

We note that the fin structure introduced in the sample stage is a key difference as compared to the temperature-controlled sample stage in traditional emissometry systems.[38] Here, we use a fin structure with one end of the fin in contact with the ceramic heater to guide the heat flow to the sample that is placed on the other end of the fin. In doing so, we can elevate the sample from the heating stage, allowing enough space for the magnetic assembly, as indicated in Fig. 3(c). A K-series thermocouple is positioned at the tip of the rod near the sample to monitor the temperature of the sample. Despite the heat loss induced by the introduction of the fin structure, we are still able to achieve sample temperatures of up to about 1000 K.

**Results and Discussion**

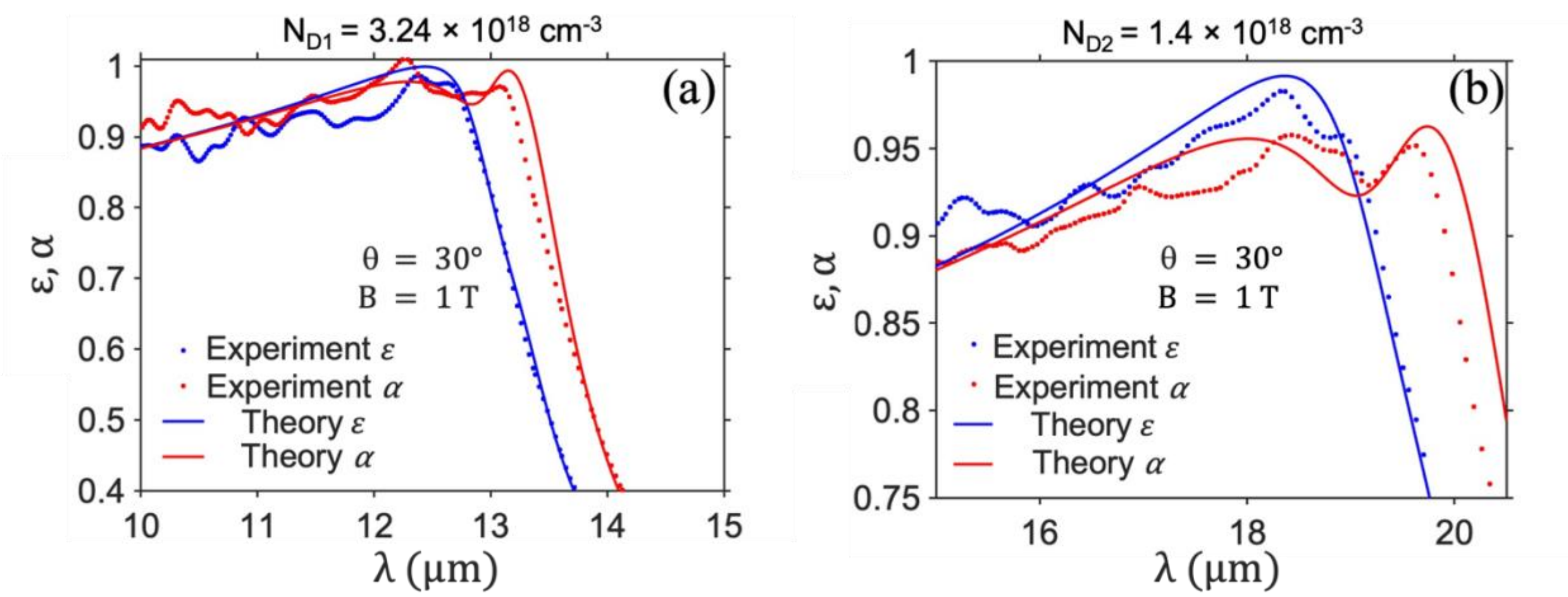

Figure 5. Radiative properties for *p*-polarized waves as a function of wavelength $\lambda$ for *n*-doped InAs samples with different doping levels at 598 K. (a) Experimental and theoretical results for the InAs sample with $N_{\mathrm{D1}} = 3.24 \times 10^{18}$ cm$^{-3}$. (b) Experimental and theoretical absorptivity results for the InAs sample with $N_{\mathrm{D2}} = 1.4 \times 10^{18}$ cm$^{-3}$.

Figures 5 (a) and (b) show the measured emissivity and absorptivity spectra for TM polarizations at $T$ = 598 K, $\theta$ = 30°, and $B$ = 1 T for two InAs samples with different doping levels of $N_{\mathrm{D1}} = 3.24 \times 10^{18}$ cm$^{-3}$ (sample 1) and $N_{\mathrm{D2}} = 1.4 \times 10^{18}$ cm$^{-3}$ (sample 2), respectively. Both samples are 525 μm thick and can be considered as opaque in the mid-infrared wavelength range.

The theoretical results based on the developed model are also overlaid, agreeing well with the experiment. For sample 2, the spectra exhibit a prominent peak around 20 μm, corresponding to the leaky Brewster mode[39,40] in the ENZ region. The Brewster mode shifts to shorter wavelengths (about 13 μm) for sample 1 due to the increase of the plasma frequency, as shown in Fig. 5(a).

The emissivity and absorptivity spectra show a significant difference, particularly near the region where the Brewster modes are excited. The difference clearly demonstrates the strong nonreciprocal response, which breaks the conventional Kirchhoff's law of thermal radiation at approximately 600 K. We note that the enhancement of nonreciprocity by Brewster modes is even stronger at larger angles as shown in Fig. S2 in the Supplementary Materials. Here we focus on investigating $\theta = 30°$ to achieve a better signal-to-noise ratio in the experiment. Thus, the nonreciprocal contrast can be greater at larger angles. As control experiments, we also conduct the same measurement for the TM polarization with $B = 0$ T and the TE polarization (electric field along the $z$ direction). In both cases, we observe overlapping emissivity and absorptivity spectra with no contrast, obeying the conventional Kirchhoff's law of thermal radiation.

As we change the temperature, the nonreciprocal properties exhibit prominent change, as shown in Fig. 6. Both samples are tested at three other different temperatures. As in Fig. 5, the nonreciprocal behavior occurs in all cases. As temperature increases, the peak caused by the Brewster mode shifts slightly to longer wavelengths, which is more clearly observed from the absorptivity spectra for both samples. Since the plasma frequency tends to increase as a function of temperature as shown in Supplementary Materials, the shift of the Brewster mode is mainly caused by the increase of the $\varepsilon_\infty$ as temperature increases, which results in the ENZ region occurring at slightly longer wavelengths.

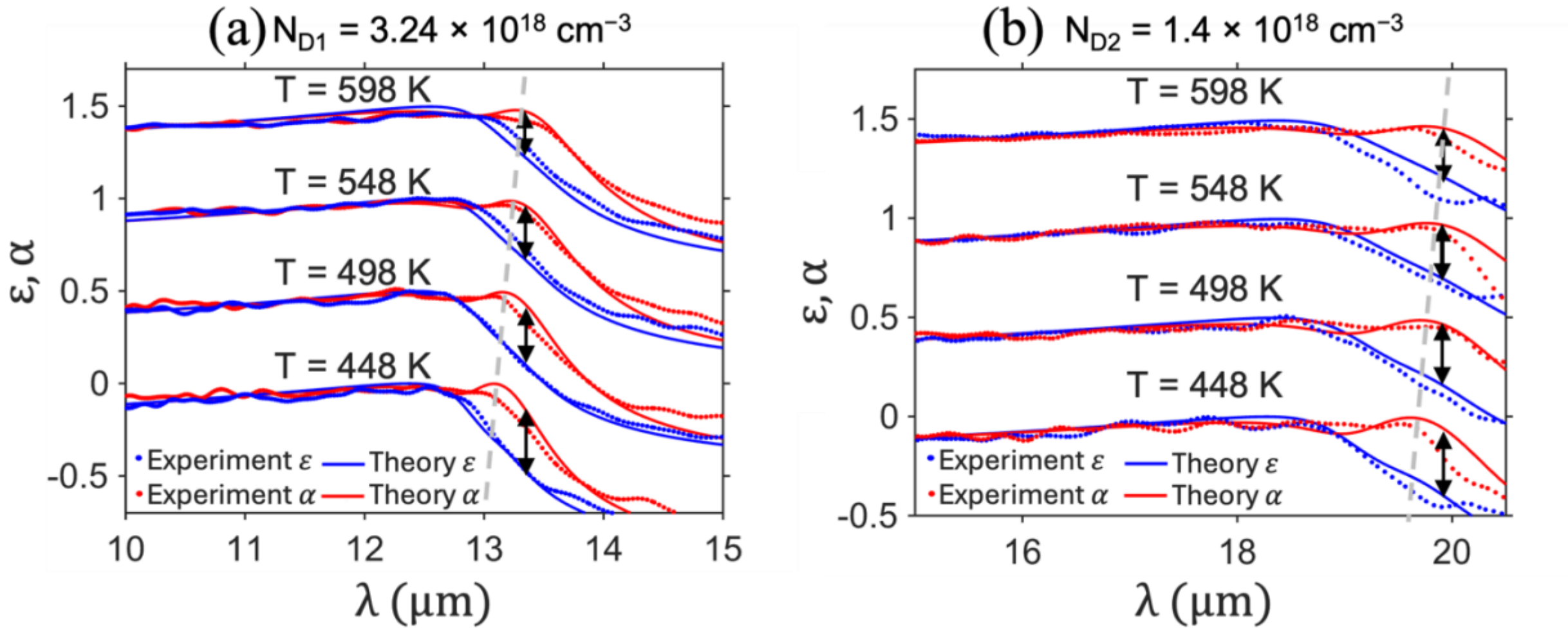

Figure 6. Absorptivity and emissivity measurements and calculations at different temperatures at $\theta$ = 30°. These results are shifted by 0.5. (a) The sample with $N_{D1} = 3.24 \times 10^{18}$ cm$^{-3}$. (b) The sample with $N_{D2} = 1.4 \times 10^{18}$ cm$^{-3}$. Arrows are added to better visualize the contrast change at the same wavelength. Dashed lines show the red shift for higher temperatures.

Besides the shift of the Brewster modes, higher temperatures also bring a strong effect on the nonreciprocal contrast. In Fig. 6, we mark the maximum contrast between $\varepsilon$ and $\alpha$ with arrows. We first determine the maximum contrast near the Brewster mode and its corresponding wavelength based on the computed spectrum at $T$ = 598 K. Then we mark the contrast at the same wavelength for all temperatures with double-headed arrows, clearly showcasing the decrease of the contrast at a fixed wavelength when temperature increases. Furthermore, dashed lines illustrate changes in the wavelength corresponding to the peak location of the Brewster modes for the absorptivity spectra. As another approach to illustrate the temperature effect on the contrast, Fig. 7 shows the maximum contrast and the corresponding wavelength as a function of the temperature for both samples. Due to the shift of the Brewster modes, the wavelength at which the maximum contrast occurs also shifts to longer wavelengths as the temperature increases. Meanwhile, the maximum contrast gradually decreases as the temperature increases. This is primarily caused by the increase in the scattering rate, which also broadens the Brewster modes. Our proposed theoretical model captures both temperature effects well, demonstrating the effectiveness of the model.

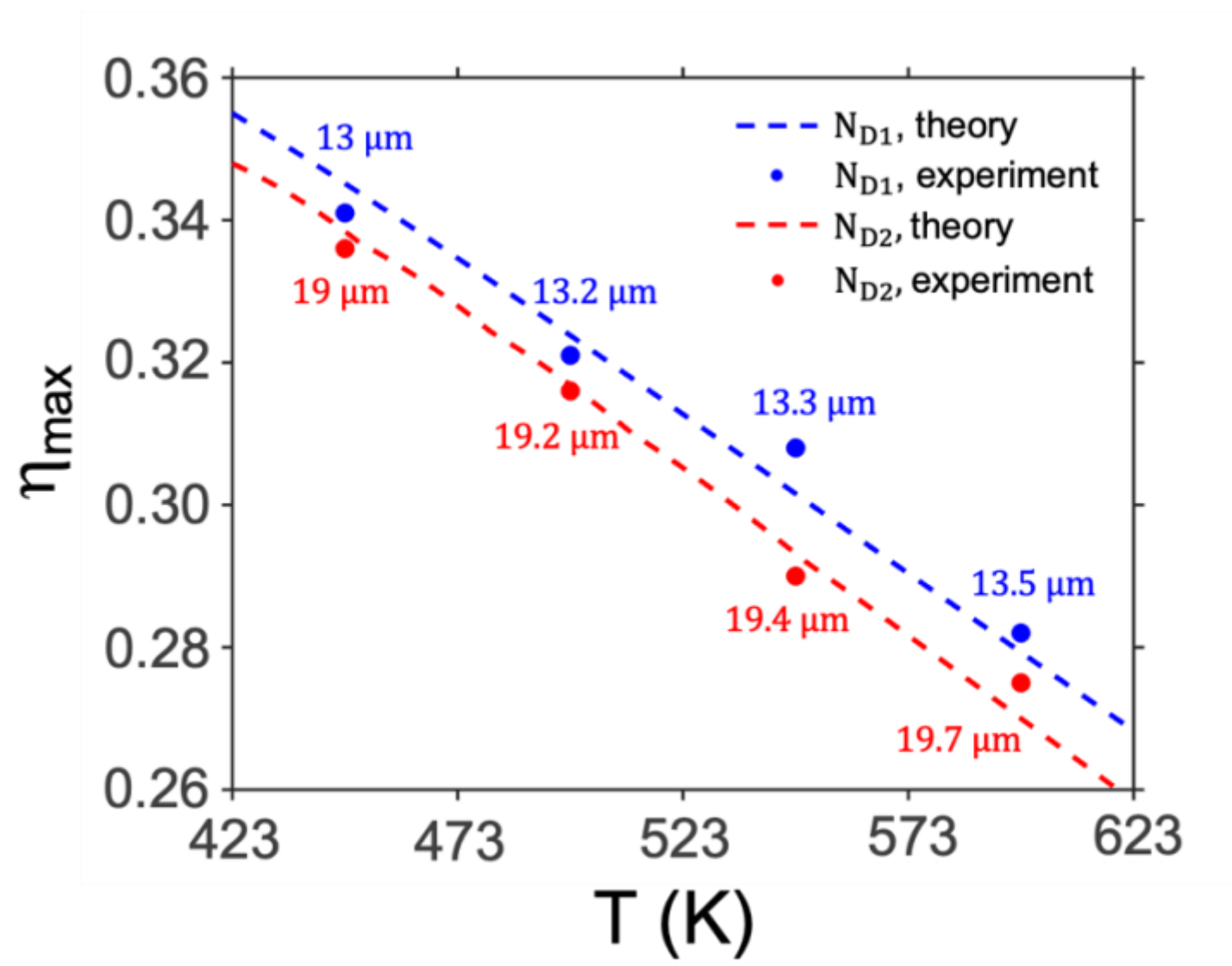

Figure 7. Maximum contrast between emissivity and absorptivity ($\eta_{\mathrm{max}}$) for *n*-doped InAs samples. $N_{\mathrm{D1}}$ and $N_{\mathrm{D2}}$ refer to the samples with carrier concentrations of $3.24 \times 10^{18}$ $\mathrm{cm}^{-3}$ and $1.4 \times 10^{18}$ $\mathrm{cm}^{-3}$, respectively. The corresponding wavelengths at which the maximum contrast occurs are marked.

The temperature-dependent nonreciprocal properties have several implications. So far, the design of nonreciprocal thermal emitters has been limited to room temperature or near room temperatures.[15,17] These designs may need to be reexamined if one would apply them to high temperature applications. This also applies to magnetic Weyl semimetals, whose nonreciprocal properties could reduce significantly as temperature increases because of the sensitive temperature response near the Weyl cones, as discussed in Ref. 38. However, for semiconductors, our results reveal that the nonreciprocal response is still much more robust than Weyl semimetals against temperature change, which opens the possibility for high-temperature nonreciprocal thermal emission control.

In our study, we limit the sample temperature to 598 K since InAs could desorb when heated up to temperatures over 703 K.[41] The desorption temperature of semiconductors typically increases as the band gap increases.[42] For example, it has been experimentally demonstrated that gallium arsenide (GaAs) can be stable even up to about 800 K.[43,44] Therefore, one could potentially utilize wider band gap materials for nonreciprocal thermal radiation control at even higher temperatures.

Motivated by these thoughts, we conduct a calculation based on our proposed theoretical model for $In_xGa_{1-x}As$ semiconductors. In Fig. 8, we show the nonreciprocal contrast for TM waves at $\theta$ = 80° and $B$ = 1 T for InAs, GaAs, and $In_{0.53}Ga_{0.47}As$, which can be readily grown on indium phosphide (InP).[45] In the calculations, we assume all materials are *n*-doped with the same doping level of $8.5 \times 10^{17}$ $cm^{-3}$. We choose a slightly lower doping level as compared to the previous InAs samples to highlight the fact that the high nonreciprocal contrast is not very sensitive to the change in doping levels. Strong nonreciprocity is observed for all material systems, with the contrast gradually decreasing with an increasing temperature. For GaAs, the contrast is not as strong as InAs due to its heavy electron mass.[35] The ENZ region also occurs at a longer wavelength range as indicated by the bright high-contrast band due to its larger band gap. The behavior of $In_{0.53}Ga_{0.47}As$ falls between InAs and GaAs in terms of wavelength and contrast values, indicating that the nonreciprocal properties can be tuned by varying $x$ in the compound.[22] The results point the pathway for ultra-high-temperature nonreciprocal thermal radiation.

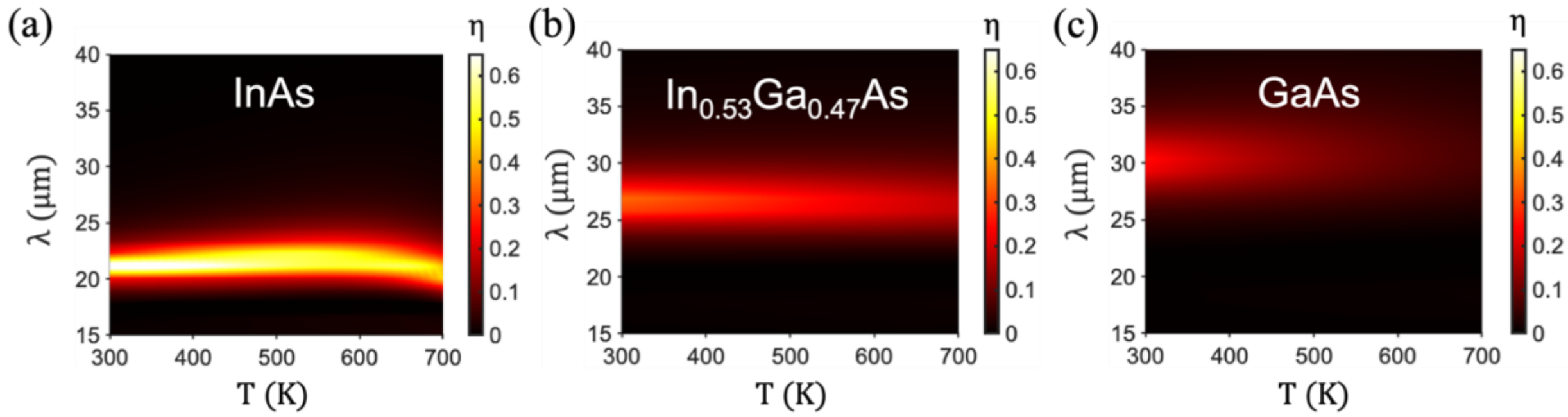

Figure 8. Contours for the nonreciprocal contrast as a function of temperature and wavelength for *n*-doped (a) InAs, (b) $In_{0.53}Ga_{0.47}As$, and (c) GaAs. All materials have the same doping level of $8.5 \times 10^{17}$ cm$^{-3}$.

While writing this manuscript, we were made aware of a work[36] that utilizes InGaAs for nonreciprocal thermal radiation at a temperature of 540 K. The results are consistent with our study, especially the results in Figs. 8(b) and S2(b) in the Supplementary Materials.

## Conclusion

In conclusion, this work demonstrates robust, high-temperature nonreciprocal thermal radiation in semiconductors, advancing the frontier of active radiative heat flow control. By developing a comprehensive theoretical model that accounts for temperature-dependent carrier dynamics, scattering rates, and permittivity shifts in magneto-optical materials, we successfully predicted and experimentally validated strong nonreciprocal contrasts in *n*-doped InAs at temperatures exceeding 600 K. Our customized angle-resolved magnetic emissometry setup revealed persistent nonreciprocity at elevated temperatures, with experimental results aligning closely with our theoretical predictions. The observed red-shifting of epsilon-near-zero regions and moderated contrast reduction at higher temperatures highlight the resilience of semiconductor-based nonreciprocal emitters under thermal stress. We believe that wide-band gap semiconductors like GaAs could sustain nonreciprocal effects beyond 800 K, unlocking applications in high-temperature radiative heat flow control and energy harvesting. This study establishes a framework for designing temperature-robust nonreciprocal systems, bridging critical gaps in transitioning lab-scale demonstrations to real-world thermal management technologies.

## ▪ ASSOCIATED CONTENT

Supplementary Materials

## ▪ AUTHOR INFORMATION

### Corresponding Author

**Bo Zhao**, *Department of Mechanical Engineering, University of Houston, Houston, TX 77204, United States*

Email: bzhao8@uh.edu

### Author

**Bardia Nabavi**, *Department of Mechanical Engineering, University of Houston, Houston, TX 77204, United States*

Email: bnabavi@uh.edu

**Sina Jafari Ghalekohneh**, *Department of Mechanical Engineering, University of Houston, Houston, TX 77204, United States*

Email: sinajafari@uh.edu

**Komron J. Shayegan**, *Department of Electrical Engineering, California Institute of Technology, Pasadena, CA, 91125, USA*

Email: shayegan@caltech.edu

**Eric J. Tervo**, *Department of Mechanical Engineering, University of Wisconsin-Madison, Madison, WI, 53706, USA*

Email: tervo@wisc.edu

**Harry Atwater,** *Department of Mechanical Engineering, University of Wisconsin-Madison, Madison, WI, 53706, USA*

Email: haa@caltech.edu

**Note**

The authors declare no competing financial interest.

## ACKNOWLEDGEMENTS

The authors acknowledge the funding from the University of Houston through the SEED program and the National Science Foundation under Grant No. CBET-2314210, and the support of the Research Computing Data Core at the University of Houston for assistance with the calculations carried out in this work. B. N. and B. Z. thank Sahag Bozoian, Aniroodh Sivaraman, and Jamar Murray for their assistance in fabricating the magnetic assembly.

## References

(1) Fan, S. Thermal photonics and energy applications. *Joule* **2017**, *1* (2), 264-273.

(2) Zhu, L.; Guo, Y.; Fan, S. Theory of many-body radiative heat transfer without the constraint of reciprocity. *Physical Review B* **2018**, *97* (9), 094302.

(3) Snyder, W. C.; Wan, Z.; Li, X. Thermodynamic constraints on reflectance reciprocity and Kirchhoff's law. *Appl. Opt.* **1998**, *37* (16).

(4) Zhao, B.; Shi, Y.; Wang, J.; Zhao, Z.; Zhao, N.; Fan, S. Near-complete violation of Kirchhoff's law of thermal radiation with a 0.3T magnetic field. *Optics Letters* **2019**, *44* (17), 4203-4206.

(5) Zhang, Z.; Zhu, L. Nonreciprocal thermal photonics for energy conversion and radiative heat transfer. *Physical Review Applied* **2022**, *18* (2), 027001.

(6) Khandekar, C.; Pick, A.; Johnson, S. G.; Rodriguez, A. W. Radiative heat transfer in nonlinear Kerr media. *Physical Review B* **2015**, *91* (11), 115406.

(7) Ghanekar, A.; Wang, J.; Fan, S.; Povinelli, M. L. Violation of Kirchhoff's law of thermal radiation with space–time modulated grating. *ACS Photonics* **2022**, *9* (4), 1157-1164.

(8) Hadad, Y.; Soric, J. C.; Alu, A. Breaking temporal symmetries for emission and absorption. *Proceedings of the National Academy of Sciences* **2016**, *113* (13), 3471-3475.

(9) Fan, L.; Guo, Y.; Papadakis, G. T.; Zhao, B.; Zhao, Z.; Buddhiraju, S.; Orenstein, M.; Fan, S. Nonreciprocal radiative heat transfer between two planar bodies. *Physical Review B* **2020**, *101* (8), 085407.

(10) Park, Y.; Zhao, B.; Fan, S. Reaching the ultimate efficiency of solar energy harvesting with a nonreciprocal multijunction solar cell. *Nano Letters* **2021**, *22* (1), 448-452.

(11) Jafari Ghalekohneh, S.; Zhao, B. Nonreciprocal solar thermophotovoltaics. *Physical Review Applied* **2022**, *18* (3), 034083.

(12) Park, Y.; Omair, Z.; Fan, S. Nonreciprocal thermophotovoltaic systems. *ACS Photonics* **2022**, *9* (12), 3943-3949.

(13) Bi, L.; Hu, J.; Jiang, P.; Kim, D. H.; Dionne, G. F.; Kimerling, L. C.; Ross, C. On-chip optical isolation in monolithically integrated non-reciprocal optical resonators. *Nature Photonics* **2011**, *5* (12), 758-762.

(14) Tanner, D. B. Optical Effects in Solids. *Cambridge University Press* **2019**.

(15) Shayegan, K. J.; Biswas, S.; Zhao, B.; Fan, S.; Atwater, H. A. Direct observation of the violation of Kirchhoff's law of thermal radiation. *Nature Photonics* **2023**, *17* (10), 891-896.

(16) Shayegan, K. J.; Hwang, J. S.; Zhao, B.; Raman, A. P.; Atwater, H. A. Broadband nonreciprocal thermal emissivity and absorptivity. *Light: Science & Applications* **2024**, *13* (1), 176.

(17) Liu, M.; Xia, S.; Wan, W.; Qin, J.; Li, H.; Zhao, C.; Bi, L.; Qiu, C.-W. Broadband mid-infrared non-reciprocal absorption using magnetized gradient epsilon-near-zero thin films. *Nature Materials* **2023**, *22* (10), 1196-1202.

(18) Do, B.; Ghalekohneh, S. J.; Adebiyi, T.; Zhao, B.; Zhang, R. Automated design of nonreciprocal thermal emitters via Bayesian optimization. *Journal of Quantitative Spectroscopy and Radiative Transfer* **2025**, *331*, 109260.

(19) Pierret, R. F. Semiconductor Device Fundamentals. *Addison-Wesley Publishing Company* **1996**.

(20) Shayegan, K. J.; Zhao, B.; Kim, Y.; Fan, S.; Atwater, H. A. Nonreciprocal infrared absorption via resonant magneto-optical coupling to InAs. *Science Advances* **2022**, *8* (18), eabm4308.

(21) Degheidy, A. R.; Elkenany, E. B.; Madkour, M. A. K.; AbuAli, A. M. Temperature dependence of phonons and related crystal properties in InAs, InP and InSb zinc-blende binary compounds. *Computational Condensed Matter* **2018**, *16*, e00308.

(22) Sotoodeh, M.; Khalid, A.; Rezazadeh, A. Empirical low-field mobility model for III-V compounds applicable in device simulation codes. *Journal of Applied Physics* **2000**, *87* (6), 2890-2900.

(23) Casimir, H. B. G. On Onsager's principle of microscopic reversibility. *Reviews of Modern Physics* **1945**, *17* (2-3), 343.

(24) Guo, C.; Zhao, B.; Fan, S. Adjoint Kirchhoff's law and general symmetry implications for all thermal emitters. *Physical Review X* **2022**, *12* (2), 021023.
(25) Yang, C.; Zhao, B.; Cai, W.; Zhang, Z. M. Mid-infrared broadband circular polarizer based on Weyl semimetals. *Optics Express* **2022**, *30* (2), 3035-3046.
(26) Zhu, L.; Fan, S. Near-complete violation of detailed balance in thermal radiation. *Physical Review B* **2014**, *90* (22), 220301.
(27) Madelung, O. Semiconductors: data handbook. *Springer Science & Business Media* **2004**.
(28) Paul, S.; Roy, J. B.; Basu, P. K. Empirical expressions for the alloy composition and temperature dependence of the band gap and intrinsic carrier density in $Ga_xIn_{1-x}As$. *Journal of Applied Physics* **1991**, *69* (2), 827-829.
(29) Zielinski, E.; Schweizer, H.; Streubel, K.; Eisele, H.; Weimann, G. Excitonic transitions and exciton damping processes in InGaAs/InP. *Journal of Applied Physics* **1986**, *59* (6), 2196-2204.
(30) Karachevtseva, M.; Ignat'ev, A.; Mokerov, V.; Nemtsev, G.; Strakhov, V.; Yaremenko, N. Temperature dependence of the photoluminescence of $In_xGa_{1-x}As$/GaAs quantum-well structures. *Semiconductors* **1994**, *28* (7), 691-694.
(31) Varshni, Y. P. Temperature dependence of the energy gap in semiconductors. *Physica D: Nonlinear Phenomena* **1967**, *34*, 149-154.
(32) Fang, Z.; Ma, K.; Jaw, D.; Cohen, R.; Stringfellow, G. Photoluminescence of InSb, InAs, and InAsSb grown by organometallic vapor phase epitaxy. *Journal of Applied Physics* **1990**, *67* (11), 7034-7039.
(33) Sze, S. M. Semiconductor devices: physics and technology. *John wiley & sons* **2008**.
(34) Bouarissa, N.; Aourag, H. Effective masses of electrons and heavy holes in InAs, InSb, GaSb, GaAs and some of their ternary compounds. *Infrared Physics and Technology* **1999**, *40*, 343-349.
(35) Nakwaski, W. Effective masses of electrons and heavy holes in GaAs, InAs, A1As and their ternary compounds. *Physica B: Condensed Matter* **1995**, *210* (1), 1-25.
(36) Zhang, Z.; Dehaghi, A. K.; Ghosh, P.; Zhu, L. Observation of Strong Nonreciprocal Thermal Emission. *arXiv preprint arXiv:2501.12947* **2025**.
(37) Adambukulam, C.; Sewani, V. K.; Stemp, H. G.; Asaad, S.; Mądzik, M. T.; Morello, A.; Laucht, A. An ultra-stable 1.5 T permanent magnet assembly for qubit experiments at cryogenic temperatures. *Review of Scientific Instruments* **2021**, *92* (8).
(38) Shan, S.; Chen, C.; Loutzenhiser, P. G.; Ranjan, D.; Zhou, Z.; Zhang, Z. M. Spectral emittance measurements of micro/nanostructures in energy conversion: a review. *Frontiers in Energy* **2020**, *14*, 482-509.
(39) Taliercio, T.; Guilengui, V. N.; Cerutti, L.; Tournié, E.; Greffet, J.-J. Brewster "mode" in highly doped semiconductor layers: an all-optical technique to monitor doping concentration. *Optics express* **2014**, *22* (20), 24294-24303.
(40) Jafari Ghalekohneh, S.; Du, C.; Zhao, B. Controlling the contrast between absorptivity and emissivity in nonreciprocal thermal emitters. *Applied Physics Letters* **2024**, *124* (10).
(41) Yamaguchi, H.; Horikoshi, Y. As desorption from GaAs and AlAs surfaces studied by improved high-energy electron reflectivity measurements. *Journal of Applied Physics* **1992**, *71* (4), 1753-1759.
(42) Laughlin, D.; Wilmsen, C. Thermal oxidation of InAs. *Thin Solid Films* **1980**, *70* (2), 325-332.
(43) Madhusoodhanan, S.; Sabbar, A.; Tran, H.; Lai, P.; Gonzalez, D.; Mantooth, A.; Yu, S.-Q.; Chen, Z. High-temperature analysis of optical coupling using AlGaAs/GaAs LEDs for high-density integrated power modules. *Scientific Reports* **2022**, *12* (1), 3168.
(44) Millea, M.; Kyser, D. Thermal decomposition of gallium arsenide. *Journal of Applied Physics* **1965**, *36* (1), 308-313.
(45) Ram, R.; Dudley, J.; Bowers, J.; Yang, L.; Carey, K.; Rosner, S.; Nauka, K. GaAs to InP wafer fusion. *Journal of Applied Physics* **1995**, *78* (6), 4227-4237.